\documentclass[aps,twocolumn,prb,amsmath,amssymb,floatfix,superscriptaddress,longbibliography]{revtex4-2}

\usepackage[utf8]{inputenc}
\usepackage{graphicx,color,xcolor}% Include figure files
\usepackage[colorlinks=true,citecolor=blue]{hyperref}
\usepackage{dcolumn}% Align table columns on decimal point
\usepackage{bm}% bold math
\usepackage{amsmath}
\usepackage{siunitx}
\usepackage{orcidlink}

\newcommand{\eps}{\varepsilon}

\newcommand{\ie}{\emph{i.\,e.}}
\newcommand{\eg}{\emph{e.\,g.}}

\newcommand{\mol}{$[\text{Fe}(\text H_2\text B(\text{pz})(\text{pypz}))_2]$}

\newcommand{\ieap}{Institut für Experimentelle und Angewandte Physik, Christian-Albrechts-Universität zu Kiel, 24098 Kiel, Germany}
\newcommand{\iic}{Institut für Anorganische Chemie, Christian-Albrechts-Universität zu Kiel, 24098 Kiel, Germany}

\begin{document}

\title{Statistical analysis of electron-induced switching of a spin-crossover complex}

\author{Jonas Fußangel\,\orcidlink{0009-0001-9567-5777}}
\affiliation{Fakultät für Physik and CENIDE, Universität Duisburg-Essen, 47048 Duisburg, Germany}
\author{Björn Sothmann\,\orcidlink{0000-0001-9696-9446}}
\affiliation{Fakultät für Physik and CENIDE, Universität Duisburg-Essen, 47048 Duisburg, Germany}
\author{Sven Johannsen}
\affiliation{\ieap}
\author{Sascha Ossinger}
\affiliation{\iic}
\author{Felix Tuczek\,\orcidlink{0000-0001-7290-9553}}
\affiliation{\iic}
\author{Richard Berndt\,\orcidlink{0000-0003-1165-9065}}
\affiliation{\ieap}
\author{Jürgen König\,\orcidlink{0000-0003-3836-4611}}
\affiliation{Fakultät für Physik and CENIDE, Universität Duisburg-Essen, 47048 Duisburg, Germany}
\author{Manuel Gruber\,\orcidlink{0000-0002-8353-4651}}
\affiliation{Fakultät für Physik and CENIDE, Universität Duisburg-Essen, 47048 Duisburg, Germany}

\date{\today}

\begin{abstract}
Spin-crossover complexes exhibit two stable configurations with distinct spin states.
The investigation of these molecules using low-temperature scanning tunneling microscopy has opened new perspectives for understanding the associated switching mechanisms at the single-molecule level.
While the role of tunneling electrons in driving the spin-state switching has been clearly evidenced, the underlying microscopic mechanism is not completely understood.
In this study, we investigate the electron-induced switching of \mol\ (pz = pyrazole, pypz = pyridylpyrazole) adsorbed on Ag(111).
The current time traces show transitions between two current levels corresponding to the two spin states.
We extract switching rates from these traces by analyzing waiting-time distributions.
Their sample-voltage dependence can be explained within a simple model in which the switching is triggered by a transient charging of the molecule.
The comparison between experimental data and theoretical modeling provides estimates for the energies of the lowest unoccupied molecular orbitals, which were so far experimentally inaccessible.
Overall, our approach offers new insights into the electron-induced switching mechanism and predicts enhanced switching rates upon electronic decoupling of the molecule from the metallic substrate, for example by introducing an ultrathin insulating layer.
\end{abstract}

\maketitle

\section{\label{sec:Introduction}Introduction}

Spin-crossover (SCO) complexes may be switched between a low-spin (LS) and a high-spin (HS) state \cite{gutlich_spin_2000}.
The switching may be triggered by temperature, light, or electrical current and affects magnetic, optical, and electronic properties of the complexes.
Therefore,
this class of compounds has received considerable attention in view of technological applications, such as data storage, smart pigments, and sensors \cite{molnar_spin_2018}.
In addition, SCO complexes give the prospect of miniaturized devices, down to single molecules, as the switching is an intrinsic property of the molecules.

Downsizing to the scale of single molecules %also 
requires an adapted, localized trigger and readout.
The change of conductance of a molecular junction upon spin crossover has been evidenced with transport measurements \cite{prins_room-temperature_2011, osorio_conductance_2010, harzmann_single-molecule_2015, frisenda_stretching-induced_2016, shalabaeva_room_2018, hao_nonvolatile_2019, zhang_resistance_2020, gee_multilevel_2020, van_geest_contactless_2020, konstantinov_electrical_2021,dayen_room_2021, stegmann_statistical_2021,torres-cavanillas_bistable_2024} and scanning tunneling microscopy (STM) \cite{gopakumar_electron-induced_2012, miyamachi_robust_2012, gruber_spin_2014, gruber_spin_2017, jasper-tonnies_deposition_2017, jasper-toennies_robust_2017, atzori_thermal_2018, kobke_reversible_2020, brandl_iron_2020, zhang_anomalous_2020, gruber_spin-crossover_2020, tong_voltage-induced_2021, johannsen_three-state_2022, johannsen_spin-crossover_2023, johannsen_electron-induced_2021, johannsen_spin_2022,johannsen_spin-state_2024}.
In most of these studies, the switching is realized by current injection under the application of a DC sample voltage between the tip and the sample. 
The switching is thus effectively caused by the transfer of energy from the charge carriers, by the associated electric field on the molecule, or by a combination of both \cite{montenegro-pohlhammer_mechanisms_2025}.
The nature of the relevant excitation most probably depends on the system.
In STM experiments, the electrical current through the SCO molecule and the sample voltage can be modified independently by adjusting the tunneling gap between STM tip and sample/molecule.
Detailed analyses of the switching rate of a single Fe$^{2+}$ SCO complex as a function of the current (at fixed sample voltage) indicated that switching is induced by individual tunnel electrons \cite{miyamachi_robust_2012,johannsen_electron-induced_2021,johannsen_spin-state_2024}.
The switching yield, \ie\ the switching probability per tunnel electron, strongly depends on the sample voltage.
This suggests that molecular orbitals play a role in the switching mechanism.
In line with this suggestion, indications of a charged state have been observed \cite{johannsen_electron-induced_2021}, and the spin-state switching of a Co complex was tentatively interpreted in terms of electron injection and removal from molecular orbitals \cite{johannsen_spin_2022}.

Up to date, many questions remain open regarding the underlying microscopic mechanism for the switching.
What is the relevance of molecular orbitals in the electron-induced switching?
How can the efficiency of the switching be improved?
If the switching is better understood, can molecular properties be extracted from the switching dynamics?

In the present study, we investigate the electron-induced spin-state switching of \mol\ on Ag(111).
Building on the STM work presented in Ref.~\cite{johannsen_electron-induced_2021}, we analyze tunneling current time traces---exhibiting fluctuations between two current states corresponding to spin-state switching---using waiting-time distributions (WTDs) to extract switching rates.
As the experimental data suggests a mechanism involving transient charging of the molecule, we develop a four-state model comprising two spin states, each of which can be either neutral or negatively charged, with transitions governed by defined rates.
The model reproduces the experimental sample-voltage dependence of the switching rates and provides estimates of the molecular orbital energies, which were not directly accessible through spectroscopy for this system.

\section{System and experimental data}\label{chap:SysExpData}
The switching behavior of individual \mol\ complexes adsorbed on Ag(111) has been thoroughly investigated in Ref.~\cite{johannsen_electron-induced_2021} as a function of tunneling current and sample voltage. With several tens of thousands of switching events recorded, this represents the most extensive STM-based statistical study of SCO switching to date. The large dataset, collected under various current and sample-voltage conditions, enables a more detailed statistical analysis of the switching dynamics, which we present in the current study.

Sublimation of \mol\ complexes onto a clean Ag(111) surface under ultrahigh vacuum conditions leads to the formation of tetramers.
STM measurements performed at approximately 5\,K revealed that, within each tetramer, two molecules located along one of the diagonals can be reversibly switched via electron injection.
In contrast, the remaining two molecules do not switch, which is attributed to their relative stacking within the tetramer, thereby structurally hindering the conformational change.

\begin{figure}
	\includegraphics[keepaspectratio,width=0.45\textwidth]{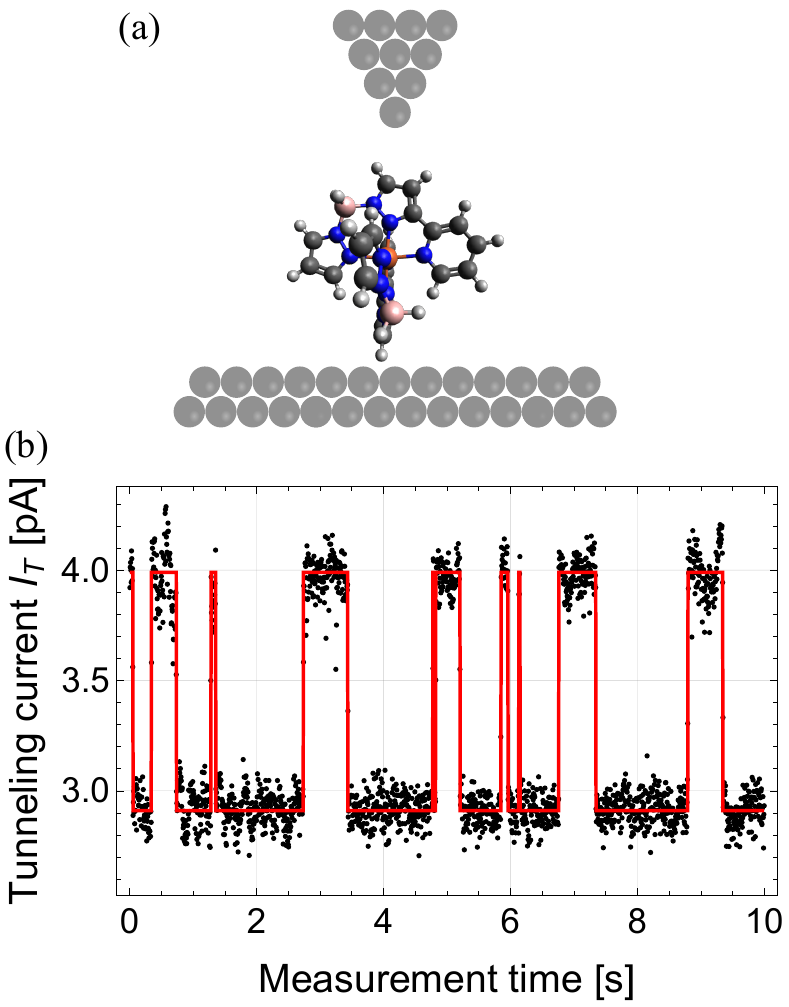}
	\caption{\label{fig:Telegraph} (a) Scheme of the experiment with a single molecule in the tunneling gap of a STM.
    (b) Exemplary time series of the tunneling current (black dots) measured across a switchable complex.
    The state with higher (lower) tunneling current represents the $H$ ($L$) state.
    This time trace was measured with a sample voltage of $\qty{1}{V}$.
    The red line is a guide to better visualize the states of the complex along with transitions between those states, for which a statistical analysis is performed.}
\end{figure}

To learn more about the switching properties, the tip was placed over a switchable molecule (Fig.~\ref{fig:Telegraph}a) and time traces of the tunneling current were measured under different tunneling conditions.
An example of such a time trace is shown in Fig.~\ref{fig:Telegraph}b.
The time trace exhibits telegraphic-like fluctuations between two current values, which are associated to the spin-state switching of the molecule under the tip. 
In the following, the spin states are denoted by $L$ and $H$, referring to the low- and high-current state, respectively.
Which of the two states carries the higher and which the lower spin value is unclear and not important for our analysis.

\begin{figure}
	\includegraphics[keepaspectratio,width=0.45\textwidth]{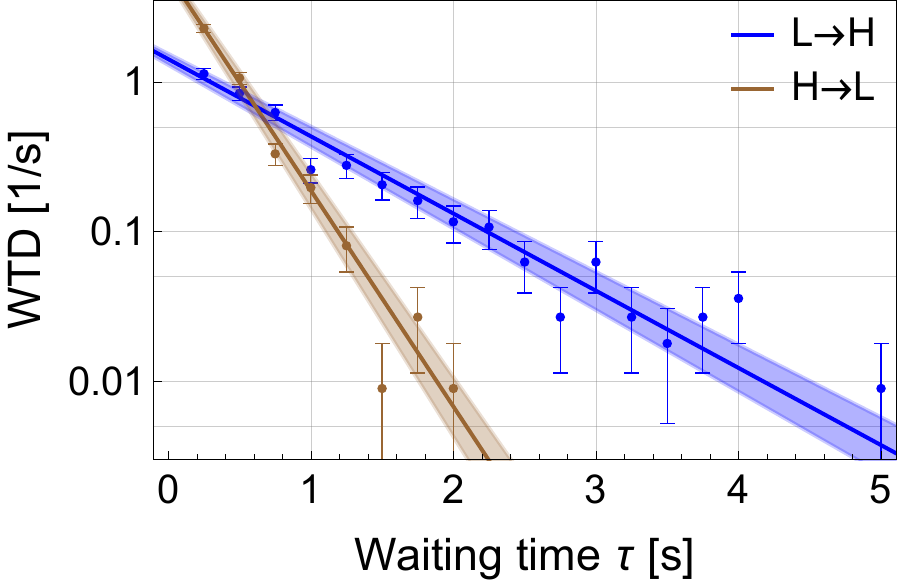}
	\caption{\label{fig:WTD} Example of waiting-time distributions (dots) using a binning of 0.25\,s displayed in a logarithmic plot. The corresponding time trace was recorded with a sample voltage of $V=\qty{1}{V}$ leading to tunneling currents of $I_{\text T,L}=\qty{2.87}{pA}$ and $I_{\text T,H}=\qty{3.90}{pA}$ in the $L$ and $H$ states, respectively.
    The solid lines are linear fits of the logarithms of the  WTD, effectively describing an exponential time decay with a decay constant of $(1.19\pm 0.07)\,\text{s}^{-1}$ for the $L\rightarrow H$    
    and $(3.33\pm 0.18)\,\text{s}^{-1}$ for the $H\rightarrow L$ transition.
    Note that the uncertainties are the ones given by the fitting routine.
    The colored areas illustrate the impact of those uncertainties.}
\end{figure}

Time traces such as shown in Fig.~\ref{fig:Telegraph}b, but spanning longer durations with a much larger number of switching events, are used to determine the waiting-time distributions (WTDs) of the switching process in the following way.
Each time the system switches into the $L$ state, the duration it remains in that state before transitioning to the $H$ state---referred to as the waiting time---is recorded.
The resulting distribution of these waiting times, using a bin size of 0.25\,s, is shown in Fig.~\ref{fig:WTD} on a logarithmic scale (blue dots).
An analogous distribution is computed for the waiting time spent in the $H$ state prior to switching back to the $L$ state (brown dots).

The WTDs exhibit an exponential decay with increasing waiting time, as expected for stochastic processes governed by a Poisson distribution.
Fitting the WTDs yields characteristic time constants whose inverses correspond to the switching rates, representing the average number of switching events per unit time.
From Fig.~\ref{fig:WTD}, we extract rates of 1.2\,s$^{-1}$ and 3.3\,s$^{-1}$ for the $L\to H$ and $H \to L$ transitions, respectively.

Determining switching rates via WTDs provides greater robustness against detection errors than the more direct procedure used in Ref.~\cite{johannsen_electron-induced_2021}, in which the number of, say, $L\to H$ switches was simply divided by the time the molecule has spent in the $L$ state.
This distinction becomes especially relevant when the switching rate approaches the bandwidth limit of the current detection, where rapid events may be missed or appear incomplete due to insufficient temporal resolution.
In such cases, the direct approach underestimates the real switching rate since undetected fast events are not taken into account.
Conversely, noise in the current can be wrongly interpreted as switching, which would lead to an overestimation of the rate (Appendix~\ref{app:WTDmethods}).
In contrast, the corresponding WTD is only affected by the detection inefficiency at short waiting times, which can be easily excluded from the exponential fit to extract the switching rates.
The use %advantage
of the WTD enables us to include additional datasets, acquired at sample voltages of 1.10 and 1.15\,V (Appendix~\ref{app:additionalData}), that were not considered in Ref.~\cite{johannsen_electron-induced_2021}.

The time series recorded at different voltages were preceded by a tip–sample adjustment to ensure comparable $I_{\text T,L}$ currents across the measurements.
To compensate for remaining variations in both $I_{\text T,L}$ and $I_{\text T,H}$, and given that the switching rate was shown to scale linearly with the tunneling current, the measured switching rates times the elementary charge were normalized by the corresponding currents.
The resulting switching yields are defined as $Y_{L\to H}:=e\tilde W_{L\to H}/I_{\text T,L}$ and the associated uncertainties are given by
\begin{align}
    \label{UncertaintyYield}\sigma_{Y_{L\to H}}=\left(\frac{\sigma_{\tilde W_{L\to H}}}{\tilde W_{L\to H}}+\frac{\sigma_{I_{\text T,L}}}{I_{\text T,L}}\right)Y_{L\to H}.
\end{align}
The average values of the currents and their corresponding standard deviations were obtained from Gaussian fits to the histograms of the time traces.

\begin{figure}
	\includegraphics[keepaspectratio,width=0.45\textwidth]{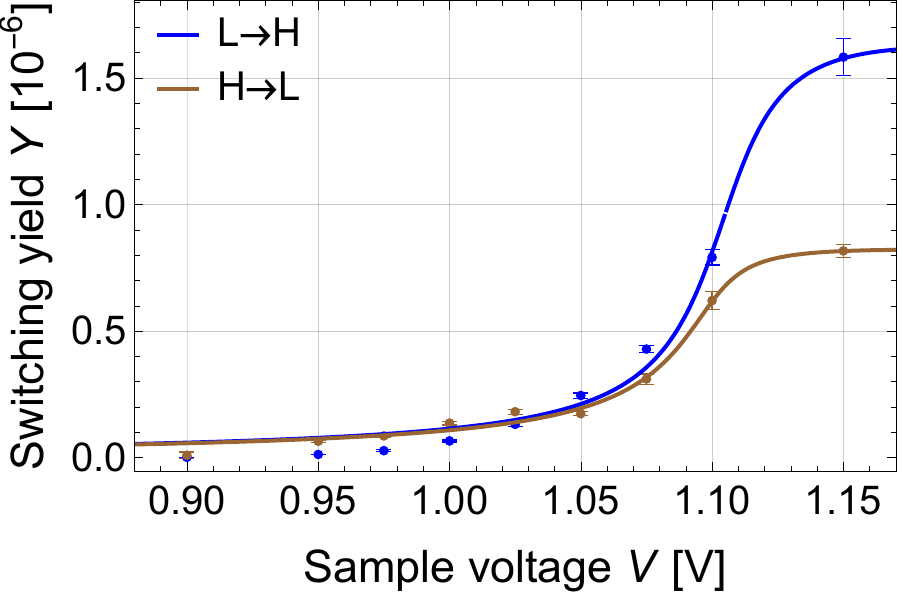}
	\caption{\label{fig:Ergebnis} Switching yields $Y_{ L\to  H}$ and $Y_{ H\to  L}$ as a function of the sample voltage $V$.
    The uncertainty bars are determined from Eq.~\eqref{UncertaintyYield}.
    The solid lines are fits of the switching yields using Eq.~\eqref{FitfunktionArctan} \cite{footnote1} for both switching directions separately, as described in section~\ref{chap:2state}.
    The fit parameters are provided in Appendix~\ref{app:fits}.}
\end{figure}

In Fig.~\ref{fig:Ergebnis}, we present the switching yields for various sample voltages ranging from 0.90 to 1.15\,V.
At 0.90\,V, the switching yield is low ($1.5\times 10^{-9}$ for the $L \rightarrow H$ direction) leading to only a few switching events in the corresponding time trace.
The yields for both switching directions, ${L \rightarrow H}$ and ${H \rightarrow L}$, gradually increase with increasing sample voltage and appear to reach a plateau near 1.15\,V.
It is this sample-voltage dependence that we are going to make use of to gain insight into the switching mechanism.

As discussed in Ref.~\cite{johannsen_electron-induced_2021}, the linear dependence between switching rate and tunneling current evidences that the spin-state switching is triggered by single-electron events.
The sharp increase of the  switching rate around a sample voltage of approximately 1\,V indicates that electron-induced switching becomes more efficient in this voltage range.
This behavior is inconsistent with mechanisms based on \textit{direct} excitation of molecular vibrations, which typically involve energies of 0.1\,eV or less.
Instead, a mechanism involving transient charging of a molecular orbital with an energy near 1\,eV is more likely.
As the sample voltage approaches the energy level of this orbital, the proportion of electrons tunneling through the molecule---and thus transiently populating the orbital---increases significantly. This enhanced charging probability results in a higher switching efficiency.
Attempts to identify the relevant molecular orbital directly via differential-conductance ($\mathrm dI/\mathrm dV$) spectroscopy have remained unsuccessful because of the switching of the molecule in the relevant sample-voltage range.
In contrast, the statistical analysis of the switching behavior presented in this paper enables us to determine the molecular-orbital energy.

\section{\label{sec:Modeling}Spin-switching model}
In the following, we describe a model based on the assumption that spin-state switching of the complexes requires transient charging.
To this end, we first determine the fraction of the tunneling current that flows through the molecular orbitals---effectively resulting in transient charging---as a function of the sample voltage.
We then compute the switching rates and yields within a four-state model that includes two spin states, each of which can exist in a neutral or negatively charged configuration, using a master-equation approach.
Finally, the model is projected onto an effective two-level system, which is, then, used for a comparison with the experimental results.

\subsection{Tunneling current through the molecule}

\begin{figure}
	\includegraphics[keepaspectratio,width=0.4\textwidth]{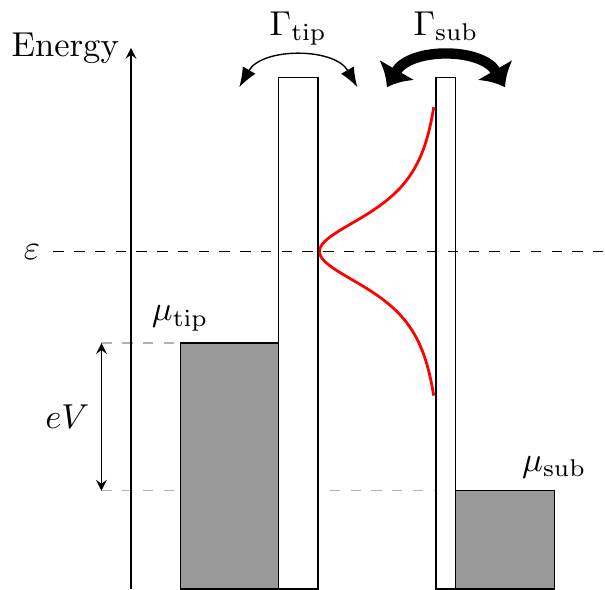}
	\caption{\label{fig:EnergieLUMO} Scheme of the electronic transport from the tip to the substrate via a molecular orbital at an energy $\epsilon$.
    The difference of the chemical potentials between the tip and the substrate is adjusted with the sample voltage $V$. The electronic coupling between the molecule and the substrate $\Gamma_\text{sub}$ is assumed to be significantly larger than that between the tip and the molecule $\Gamma_\text{tip}$, a general situation for STM experiments.
    The transmission through the molecular state (depicted in red) has a 
    %Lorentzian
    Breit-Wigner form with the width essentially determined by the electronic coupling to the substrate $\Gamma_\text{sub}$.}
\end{figure}

In both spin states, the total tunneling current ${I_\text{T}=I_\text{mol}+I_\text{dir}}$ consists of two contributions, namely the current through the molecular orbital $I_\text{mol}$ and the current that passes from the tip directly into the substrate, $I_\text{dir}$.
As the molecule is strongly coupled to the substrate, it is very unlikely that the molecular orbital is charged with more than one extra electron.
Therefore, Coulomb-blockade effects in the transport through the molecule can be neglected and transport is well described as resonant tunneling through a spin-degenerate single molecular orbital within the framework of Landauer-Büttiker theory of non-interacting electrons. The current through the molecule is then given by~\cite{buttiker_four-terminal_1986,meir_landauer_1992}
\begin{equation}
\label{I_mol_1}
	I_\text{mol}=\frac{2e}{h}\int d\omega \,\mathcal T(\omega)\left[f_\text{tip}(\omega)-f_\text{sub}(\omega)\right],
\end{equation}
where $f_\text{tip}(\omega)$ and $f_\text{sub}(\omega)$ denote the Fermi distribution of electrons in the tip and substrate, respectively. The transmission function $\mathcal T(\omega)$ is of Breit-Wigner form
\begin{equation}
\label{Transmission}
	\mathcal T(\omega)=\frac{\Gamma_\text{sub}\Gamma_\text{tip}}{(\omega-\varepsilon)^2+\left(\frac{\Gamma_\text{sub}+\Gamma_\text{tip}}{2}\right)^2},
\end{equation}
with a resonance at the energy $\varepsilon$ of the lowest-unoccupied molecular orbital (LUMO).
The width of the resonance is determined by the sum of the electronic coupling between the molecule and the substrate, $\Gamma_\text{sub}$, and the coupling between the molecule and the tip, $\Gamma_\text{tip}$.
These couplings are taken to be energy independent, which assumes a constant density of states of the tip and substrate as well as constant tunnel amplitudes.
A scheme of the transport through a molecular orbital is shown in Fig.~\ref{fig:EnergieLUMO}.

The expression for the current through the molecule can be simplified for the regime in which the experiment is performed.
First, the temperature of \qty{5}{K} is so low that the Fermi functions can be  approximated by step functions, such that the integral in Eq.~(\ref{I_mol_1}) can be performed analytically. 
In addition, due to the strong asymmetry $\Gamma_\text{sub}\gg \Gamma_\text{tip}$ in the couplings of the molecular orbital to the substrate and the tip, the width of the Breit-Wigner resonance is determined by $\Gamma_\text{sub}$ alone.
The strong coupling asymmetry, furthermore, implies that the sample voltage $V$ applied between tip and substrate mainly drops across the tunnel barrier between tip and molecule, \ie\ the sample voltage $V$ enters Eq.~(\ref{I_mol_1}) only through $f_\text{tip}(\omega)$ \footnote{It is worth noting that a voltage drop between the molecule and the metal substrate has been reported in previous studies (\eg, Refs.~\cite{karan_shifting_2015,johannsen_three-state_2022}).
For the sake of simplicity, however, we choose to neglect this voltage drop in our model.
This approximation facilitates the analysis but leads to a potential overestimation of the orbital energies.}.
As a result, Eq.~(\ref{I_mol_1}) simplifies to the compact analytical expression
\begin{equation}
\label{CurrentMol}
	I_\text{mol}=\frac{4e \Gamma_\text{tip}}{h}\left[\arctan \frac{2\varepsilon}{\Gamma_\text{sub}}-\arctan\frac{2(\varepsilon-eV)}{\Gamma_\text{sub}}\right].
\end{equation}
The maximal current $2e\Gamma_\text{tip}/\hbar$ through the molecule would be achieved if the molecular state was fully confined within the energy window between the chemical potentials of tip and substrate.
In our experiment, however, the molecular orbital lies partially outside this window and the current is accordingly reduced. 
Nevertheless, we use $e\Gamma_\text{tip}/\hbar$ as the current scale with which the direct current from tip to substrate can be compared.
This motivates us to write the direct current in the form
\begin{equation}
\label{DirectCurrent}
	I_\text{dir}=\frac{e\Gamma_\text{tip}}{\hbar} \alpha V \, .
\end{equation}
It is proportional to the sample voltage $V$.
The factor $\alpha$ includes the density of states of the substrate.
In addition, it accounts for the fact that the tunneling distance from tip to substrate is larger than from tip to molecule, which results in a reduced tunneling amplitude.

\begin{figure}
	\includegraphics[width=\columnwidth]{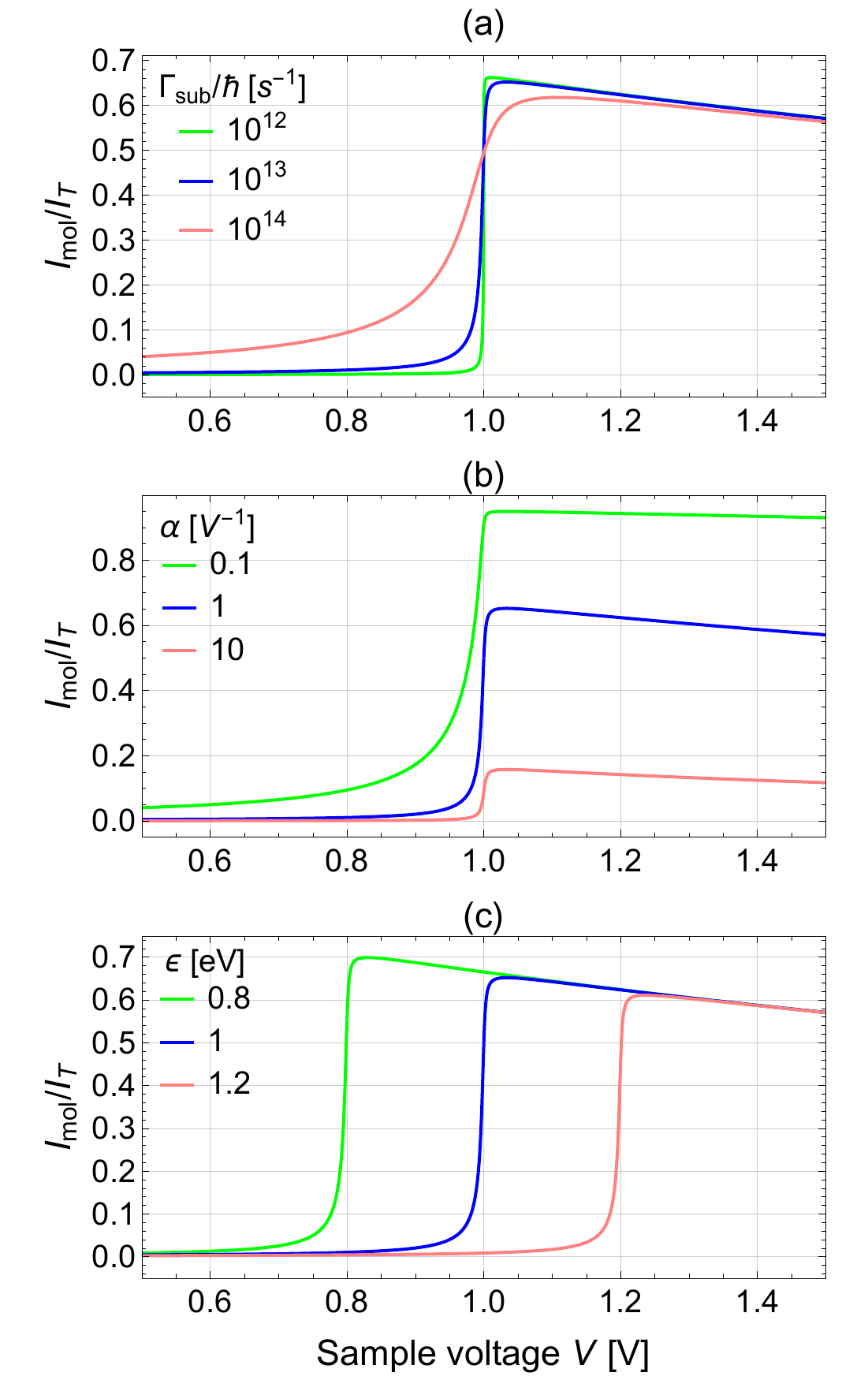}
	\caption{\label{fig:directCurrent} Fraction of the current via the molecular orbital $I_{\text{mol}}/I_\text{T}$ as a function of the sample voltage $V$ for different parameters (a) $\Gamma_{\text{sub}}/\hbar$, (b) $\alpha$,  and (c) $\eps$.
    Note that these plots assume a constant-current measurement mode, where the tip-substrate distance is adjusted to maintain $I_T$ constant. In each plot all constant parameters are set to $\Gamma_{\text{sub}}/\hbar = 10^{13}$\,s$^{-1}$, $\alpha = 1$\,V$^{-1}$, and $\eps = 1$\,eV which corresponds to the blue curves.}
\end{figure}

The spin-switching yields are measured for different sample voltages $V$ while keeping the total tunneling current $I_\text{T}$ constant.
This is achieved by adapting the distance between the STM tip and the molecule, which affects both the tip-molecule coupling $\Gamma_\text{tip}$ and the tip-substrate conductance $\text{d} I_\text{dir}/\text{d} V=e\Gamma_\text{tip} \alpha/\hbar$.
In lack of more detailed microscopic information, we assume that the \textit{ratio} of tunneling amplitudes from the tip to the molecule and the substrate is only weakly affected by the adjustment of the tip position, such that $\alpha$ is approximately voltage independent. 
We can, then, solve Eq.~(\ref{DirectCurrent}) for $\Gamma_\text{tip}$, plug this into Eq.~(\ref{CurrentMol}) and use $I_\text{dir}=I_\text{T}-I_\text{mol}$ to express the voltage-dependent fraction
\begin{equation}
\label{ImolGesamt}
	\frac{I_\text{mol}}{I_\text{T}}=\left(1+\frac{\pi}{2}\frac{\alpha V}{\arctan \frac{2\varepsilon}{\Gamma_\text{sub}}-\arctan \frac{2(\varepsilon-e V)}{\Gamma_\text{sub}}}\right)^{-1}
\end{equation}
of the total tunneling current that flows through the molecule in terms of the parameters $\Gamma_{\text{sub}}$, $\alpha$, and $\eps$. 
This is illustrated in Fig.~\ref{fig:directCurrent}.

As an exemplary case, we choose the parameter set $\Gamma_{\text{sub}}/\hbar = 10^{13}$\,s$^{-1}$, $\alpha = 1$\,V$^{-1}$, and $\eps = 1$\,eV (blue curves in all three panels of Fig.~\ref{fig:directCurrent}).
The transmission through the molecule, as described by Eq.~(\ref{Transmission}) together with $\Gamma_{\text{sub}}\gg \Gamma_\text{tip}$, becomes resonant at the energy $\eps = 1$\,eV of the molecular orbital with a full width at half maximum of $\Gamma_{\text{sub}}= \qty{6.6}{meV}$.
This results in a strongly suppressed current for small voltages and a step (here with width $\qty{6.6}{mV}$) once the position of the molecular orbital is reached (here at $1$\,V).
For voltages beyond this threshold, the ratio $I_\text{mol}/I_\text{T}$ decreases again since the direct current from tip to substrate scales linearly with $V$ even when the current through the molecule is already saturated.
As a result, there is a maximum of $I_\text{mol}/I_\text{T}$ (here with a value of around $0.65$).

As shown in panel (a) of Fig.~\ref{fig:directCurrent}, the width of the step becomes broader for larger couplings of the molecular orbital to the substrate.
This, in turn, reduces that maximum value of $I_\text{mol}/I_\text{T}$.
The dependence of $I_\text{mol}/I_\text{T}$ on $\alpha$ is depicted in panel (b).
An increased $\alpha$ corresponds to an increased direct current while the current through the molecule remains unaffected. 
This reduces the maximum value of $I_\text{mol}/I_\text{T}$.
Finally, panel (c) illustrates the role of the orbital energy $\varepsilon$.
If $\varepsilon$ is decreased then the step is shifted towards a smaller voltage, which in turn leads to an increase of the maximum value of $I_\text{mol}/I_\text{T}$, since the direct current scales linearly with the voltage.

The above analysis provides recipes to obtain the largest $I_\text{mol}/I_\text{T}$ ratio.
The electronic coupling between the molecule and the substrate as well as between the tip and the substrate should be as low a possible.
This may, for instance, be achieved by employing a thin insulating layer between the metal substrate and the molecule.
In addition, the molecular resonance should be close to the Fermi level, which can potentially be achieved by chemical engineering of the molecule's electronic states.

The energy of the molecular orbital is typically determined in STM experiments by recording differential conductance as a function of sample voltage at a fixed tip–sample distance.
In the experiment reported here, however, differential-conductance spectra in the relevant voltage range (around 1\,V) could not be reliably obtained due to the rapid back-and-forth switching of the molecule.
As a result, the molecular orbital energy could not be directly inferred experimentally.
To nevertheless gain insight into the system and the switching mechanism, we fit the measured switching data to a minimal model, which we develop in the following.

\subsection{Four-state model}
\begin{figure}[h]
	\includegraphics[width=0.9\columnwidth]{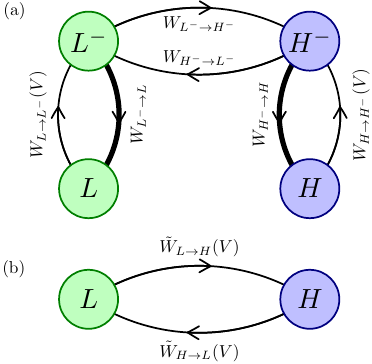}
	\caption{\label{fig:model} (a) Sketch of the four-state model with the corresponding transition rates. $L$ and $L^-$ denote the charge-neutral and negatively charged low-current state, while $H$ and $H^-$ refer to the neutral and charged high-current state. $W_{L\to L^-}$ and $W_{H\to H^-}$ are the charging rates of the low- and high-current states, $W_{L^-\to L}$ and $W_{H^-\to H}$ are the discharging rates, respectively. $W_{L^-\to H^-}$ and $W_{H^-\to L^-}$ represent the spin-state switching rates, which in our model occurs between the charged states while the direct switching between the states $L$ and $H$ is prohibited.
	(b) Sketch of the effective two-state model with the corresponding switching rates $\tilde W_{L\to H}$ and $\tilde W_{H\to L}$.
	The tunneling of electrons that gives rise to the switching is accounted for only indirectly via the calculation of the effective rates for the two-state model.}
\end{figure}

To describe the switching dynamics, we suggest the four-state model shown in Fig.~\ref{fig:model}a. 
The states $L$ and $H$ correspond to the two different, uncharged spin states of the SCO molecule resulting in a low or high current, respectively.
The current and voltage dependence of the spin-switching rate suggests that the molecule becomes transiently charged.
Therefore, we include two additional states $L^-$ and $H^-$, which correspond to the molecule charged with one excess electron relative to the uncharged $L$ and $H$ states, and we assume that spin switching occurs only between these charged states.
As a result, there are six non-vanishing transition rates between the four states.
The fact that the charged states have not been directly observed in the experiment suggests that they have a short lifetime due to a strong molecule-substrate coupling $\Gamma_{\text{sub}}$. 
Therefore, we assume that the transition rates $W_{L^-\to L}$ and $W_{H^-\to H}$ from the charged to the uncharged states are much larger than all other transition rates of the model.
This is indicated by thick lines in Fig.~\ref{fig:model}a.
Additionally, we interpret that the transition rates are $W_{L^-\to L}=\Gamma_{\text{sub},L}/\hbar$ and $W_{H^-\to H}=\Gamma_{\text{sub},H}/\hbar$.
These rates are, furthermore, assumed to be independent of the sample voltage.
This is justified by the fact that the molecular orbital (including its finite width) lies well above the Fermi level of the substrate, such that an electron tunneling out of the molecule is not hindered by Pauli blocking due to occupied states in the substrate.

The switching between the charged states $L^-$ and $H^-$ is an internal process of the molecule. 
The corresponding rate depends neither on the tunneling current $I_\text{T}$ nor the sample voltage $V$. 
Therefore, the only rates that depend on the sample voltage are the rates $W_{L\to L^-}$ and $W_{H\to H^-}$ for charging the molecule.

The dynamics of the system is governed by a master equation for the probabilities $P_L$, $P_{L^-}$, $P_{H^-}$, and $P_H$ to find the molecule in the state $L$, $L^-$, $H^-$, and $H$, respectively.
In the stationary limit, it reads
\begin{widetext}
	\begin{equation}
		\label{Ratengleichung4state}
        0
        =\frac{\mathrm d}{\mathrm dt}
        \left(\begin{array}{c} P_L \\ P_{L^-} \\ P_{H^-} \\ P_H \end{array} \right)
        =
        \left(\begin{array}{cccc}-W_{L\to L^-} & W_{L^-\to L} & 0 & 0 \\ W_{L\to L^-} & -W_{L^-\to L}-W_{L^- \to H^-} & W_{H^-\to L^-} & 0 \\0 & W_{L^-\to H^-} & -W_{H^-\to H} -W_{H^-\to L^-}& W_{H\to H^-} \\ 0 & 0 & W_{H^-\to H} & -W_{H\to H^-}\end{array}\right)
        \left(\begin{array}{c} P_L \\ P_{L^-} \\ P_{H^-} \\ P_H \end{array} \right).
	\end{equation}
\end{widetext}
The off-diagonal elements of the matrix are the transition rates between the states.
The diagonal matrix elements are fixed by the condition that each column has sum zero, which ensures that the normalization condition $P_L+ P_{L^-}+P_{H^-}+P_H=1$ is fulfilled for all times.

\subsection{Projection to an effective two-state model}\label{chap:2state}

The charged states $L^-$ and $H^-$ are extremely short lived such that they cannot be resolved in the experiment.
They only appear as intermediate states to facilitate the switching between $L$ and $H$, visible in Fig.~\ref{fig:Telegraph}b.
We, therefore, project the four-state model introduced above onto an effective two-state model that only contains the uncharged states $L$ and $H$ (Fig.~\ref{fig:model}b).

We use the first row of Eq.~(\ref{Ratengleichung4state}) to express $P_{L^-}$ in terms of $P_L$ and the fourth row to write $P_{H^-}$ in terms of $P_H$.
Plugging this into the second or third row yields
\begin{equation}
\label{master_two_state}
    P_L \frac{W_{L\to L^-}W_{L^-\to H^-}}{W_{L^-\to L}}
    = P_H \frac{W_{H\to H^-}W_{H^-\to L^-}}{W_{H^-\to H}} \, ,
\end{equation}
which has the structure $\tilde{P}_L \tilde{W}_{L\to H} = \tilde{P}_H \tilde{W}_{H\to L}$ of a stationary master equation of a two-state system.
For a consistent two-state model, we have to redistribute the probabilities $P_{L^-}$ and $P_{H^-}$ of the charged stated to $\tilde{P}_L$ and $\tilde{P}_H$.
However, since the probabilities of the charged states are much smaller than the uncharged ones, we can simply put $\tilde{P}_L = P_L$ and $\tilde{P}_H = P_H$, which approximately fulfills the normalization condition $\tilde{P}_L+\tilde{P}_H=1$.
The effective rates of the two-state model can be read off from Eq.~(\ref{master_two_state}) to be
\begin{align}
	\label{EffectiveRate3}\tilde W_{L\to H}(V)&= \frac{W_{L\to L^-}(V)W_{L^-\to H^-}}{W_{L^-\to L}},\\
	\label{EffectiveRate4}\tilde W_{H\to L}(V)&= \frac{W_{H\to H^-}(V)W_{H^-\to L^-}}{W_{H^-\to H}}.
\end{align}
As indicated, these effective rates are voltage dependent only through the rates $W_{L\to L^-}$ and $W_{H\to H^-}$ for charging the molecule.

\subsection{Waiting-time distribution (WTD)}

The time $\tau$ that the molecule spends in state $L$ before it switches back to $H$ is denoted as the waiting time.
For a two-state model, the distribution of waiting times follows a simple exponential law \cite{brandes_waiting_2008}
\begin{align}
	\label{ExponentialDistribution1}w_{{L}}(\tau)&=\tilde W_{L\to H}\exp\left(-\tilde W_{L\to H}\tau\right)
\end{align}
with the time constant given by the inverse of the rate $\tilde W_{L\to H}$.
The probability density is normalized such that the condition $\int_0^\infty w_L(\tau) d\tau =1$ is fulfilled. The WTD $w_{H}(\tau)$ for state $H$ is simply obtained by exchanging the labels $L \leftrightarrow H$.
The exponential behavior of both WTDs is clearly observed experimentally, see Fig.~\ref{fig:WTD}. It should be noted that for the fits of the WTDs (\eg \ Fig.~\ref{fig:WTD}) the normalization was not forced by the modeling function in order to get more precise values and uncertainties for the slopes interpreted as switching rates. 

In the following, we aim to express Eqs.~\eqref{EffectiveRate3} and \eqref{EffectiveRate4} in terms of system parameters and to convert the rates into yields.
Recalling that $W_{L^-\to L} \gg W_{L\to L^-}$, the current through the molecule in the $L$ state is effectively limited by the slower charging rate and can thus be written as $I_{\text{mol},L} = e W_{L\to L^-}$.
Substituting this expression into Eq.~\eqref{EffectiveRate3} yields:
\begin{align}
    \label{EffectiveRate5}\tilde W_{L\to H}(V)&= \frac{I_{\text{mol},L}(V)}{e}\frac{W_{L^-\to H^-}}{W_{L^-\to L}}.
\end{align}
The corresponding switching yield is then obtained by dividing by $I_{\text T,L}/e$:
\begin{widetext}
	\begin{equation}
		\begin{aligned}
			\label{FitfunktionArctan} Y_{{L}\to H}(V)=\frac{I_{\text{mol},L}(V)}{I_{\text T,L}}\frac{W_{L^-\to H^-}}{W_{L^-\to L}}=\frac{W_{L^-\to H^-}}{W_{L^-\to L}}\left(1+\frac{\pi}{2}\frac{\alpha V}{\arctan\left(\frac{2\eps_L}{\hbar W_{L^-\to L}}\right)-\arctan\left(\frac{2(\eps_L-eV)}{\hbar W_{L^-\to L}}\right)}\right)^{-1}
		\end{aligned}
	\end{equation}
\end{widetext}
for the switching yield in the direction $L\to H$.
%from $L$ to $H$.
The opposite yield $Y_{{H}\to L}$ is obtained by exchanging $L\leftrightarrow H$.

\section{Comparison of experimental data with the model}\label{Kap:DataAnalysis}
The experimentally determined switching yields as a function of the sample voltage $V$ (Fig.~\ref{fig:Ergebnis}) are fitted with Eq.~\eqref{FitfunktionArctan}.
The resulting functions, for $L \to H$ and $H \to L$ switching directions, reproduce to a great extent the experimental data (solid lines in Fig.~\ref{fig:Ergebnis}).
The fit parameters are listed
%described 
in Appendix~\ref{app:fits}.
Owing to the large number of parameters, the associated uncertainties are too large to provide further insights into the different rates of the model.
The situation is more favorable for the LUMO energies extracted from the fits:
\begin{equation}
	\begin{aligned}
		\eps_L=(1.105\pm 0.020) \, \text{eV}, \\
		\eps_H=(1.100\pm 0.020) \, \text{eV}.
	\end{aligned}
	\label{EnergienLH}
\end{equation}
These are in line with the expectation and we observe that the values are very close for the two current states.

\begin{figure}
	\includegraphics[keepaspectratio,width=0.45\textwidth]{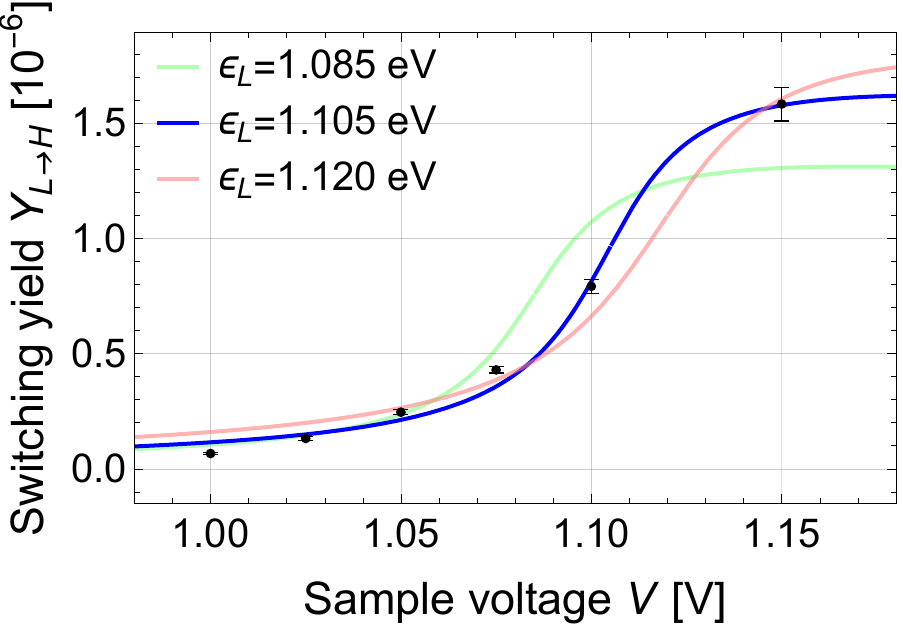}
	\par\vspace{0.5cm}
	\includegraphics[keepaspectratio,width=0.45\textwidth]{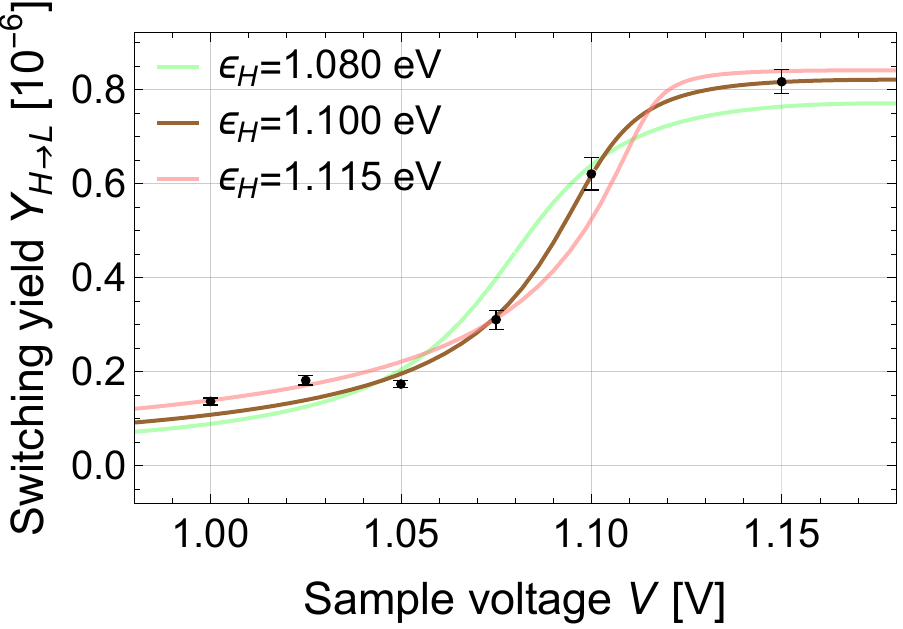}
	\caption{\label{fig:Error} Fit of the $L \to H$ (upper plot) and $H \to L$ (lower plot) switching yields inferred from the experimental data using Eq.~\eqref{FitfunktionArctan}.
    In contrast to Fig.~\ref{fig:Ergebnis}, $\eps_L$ and $\eps_H$ are fixed with the values displayed in the legend.
    The other parameters of Eq.~\eqref{FitfunktionArctan} are adjusted.}
\end{figure}

To further assess the robustness of these extracted values, we performed further fits by fixing the LUMO energy and leaving the other parameters adjustable.
The results of these additional fits are shown in Fig.~\ref{fig:Error}.
The central impact of the LUMO energy is to shift the step of the switching rate to higher or lower sample voltages. 
Focusing on the $L \to H$ switching direction (upper panel in Fig.~\ref{fig:Error}), a change of $\epsilon_L$ by $\pm 0.020$\,eV leads to an obvious deviation from the best fit of 1.105\,eV.

Density functional theory calculations of the molecular system on Ag(111)~\cite{johannsen_electron-induced_2021} place the LUMO at approximately 0.7\,eV for both the HS and LS states.
An additional unoccupied orbital is found at around 1.15\,eV for both spin states.
Our estimated orbital energies are therefore in reasonably good agreement with those inferred from the \textit{ab initio} calculations.
Assuming that the calculated LUMOs do not participate in the transport process, owing for instance to a faster orbital decay in the direction of the tip \cite{montenegro-pohlhammer_mechanisms_2025}, the agreement becomes even better.
In this scenario, the molecular switching would involve transient charging of the LUMO+1, which remains fully consistent with our model.

The developed model also provides predictions regarding the effective spin-state switching rate.
It is instructive to look back at Eq.~\eqref{FitfunktionArctan} essentially describing that the switching yield is proportional to the current through the molecule $I_\text{mol}$ and inversely proportional to the discharging rate $W_{L^- \to L}$.
Decreasing the electronic coupling between the molecule and the substrate ($\Gamma_{\text{sub},L}/\hbar = W_{L^-\to L}$) affects both variables (see also Eq.~\eqref{ImolGesamt}).

We therefore anticipate that larger spin-state switching rates can be achieved by decoupling the molecule from the metal substrate by using a thin insulating layer.
In turn, we foresee that experimental data for varying $\Gamma_\text{sub}$, \eg\ with a variable thickness of the insulating layer, and in particular determining the switching rate as a function of the insulator thickness would significantly decrease the uncertainty of the intrinsic switching rate $W_{L^-\to H^-}$, which may then become accessible.

\section{Conclusion}
We investigated electron-induced spin-state switching of \mol\ adsorbed on Ag(111), extending the analysis of the STM data presented in Ref.~\cite{johannsen_electron-induced_2021}.
In particular, we computed waiting-time distributions to extract switching rates with greater resilience against missed or false switching events, compared to direct rate determination from time traces.
The switching rates in both directions show a pronounced increase around a sample voltage of approximately 1\,V, suggesting a mechanism involving the transient population of a molecular orbital prior to spin-state transition.

Motivated by these observations, we developed a four-state model comprising two spin states---each of which can be neutral or negatively charged---and connected transitions between these states using a master-equation approach.
The molecular charging rate is assumed to scale with the fraction of the tunneling current passing through molecular orbitals, in competition with the direct tunneling into the substrate.
When projected onto an effective two-state system, the model successfully reproduces the experimentally observed evolution of the switching yields with the sample voltage.

Although some model parameters, such as the intrinsic spin-state switching rate, remain subject to large uncertainties, the model yields reliable estimates for the energies of the involved molecular orbitals---values that could not be directly accessed experimentally.

The proposed model is not limited to this specific system and can be applied to other SCO complexes on different substrates investigated via STM.
It further predicts that significantly higher switching rates could be achieved by electronically decoupling the molecule from the substrate, for instance, using an ultrathin insulating layer.
Conversely, acquiring experimental data under varying degrees of electronic coupling would help reduce uncertainties in the model parameters and may allow for a more accurate determination of the intrinsic spin-state switching rate of the SCO complex.

\acknowledgments
We acknowledge financial support from Deutsche Forschungsgemeinschaft (DFG, German Research Foundation) Projects No. 278162697-SFB 1242, 560970311, 560789750, and TU58/18-1 as well as from the Mercator Research Center Ruhr (MERCUR) within the project No.~Ko-2022-0013.
\appendix

\section{Determination of switching rates}
\label{app:WTDmethods}
In Ref.~\cite{johannsen_electron-induced_2021}, the switching rates were determined by $W_{L\to H}=N_{L\to H} / T_L$, where $N_{L\to H}$ is the number of $L\to H$ switching events in the given time series, and $T_L$ is the cumulated dwell time in the $L$ state.
The associated uncertainty is given by $\sqrt{N_{L\to H}}$.
An analogous expression is used for $W_{H\to L}$.

This method, hereafter referred to as direct counting method, for determining the switching rate is highly sensitive to spurious events (\eg\ noise falsely interpreted as switching) as well as to missed transitions.
Additionally, because the waiting times between switching events exhibit an intrinsically broad distribution, the estimated rate is strongly influenced by the sample size.
In short time series, long waiting times---though physically relevant but rare---are likely to be underrepresented or entirely absent, which can lead to an overestimation of the rate.

\begin{table}[h!]
    \centering
    \begin{tabular}{c|c|c|c}
         $V$ [V]&Method & Rate [s$^{-1}$] & Uncer. [s$^{-1}$] \\
         \hline
         &Direct counting &  $1.27$ & $0.06$ \\
        1.00&Partitioned direct counting & 0.69--1.89 & $0.28$ \\
        &WTD & $1.19$ & $0.07$ \\
        \hline
         &Direct counting &  $0.27$ & $0.02$ \\
       0.95& Partitioned direct counting & 0.08--0.60 & $0.13$ \\
       & WTD & $0.23$ & $0.02$ \\
    \end{tabular}
    \caption{Switching rate $W_{L\to H}$ at $V=1.00\,\text V$ (upper block) and $V=0.95\,\text V$ (lower block) determined with different methods.
    The segments used for the direct counting and WTD methods contain 444 and 204 switching events, respectively.
    For the partitioned method, each segment was further divided into 30 smaller sub-segments.}
    \label{tab:rateMethods}
\end{table}

The tendency of the direct counting method to overestimate switching rates in small data samples is confirmed in Tab.~\ref{tab:rateMethods}.
Applying the direct counting method to the full time series at 1.00 and 0.95\,V yields rates of 1.27 and 0.27\,s$^{-1}$, respectively.
To assess the impact of limited sampling, these time series were divided into smaller segments---segment size divided by 30 for both 1.00\,V and 0.95\,V---and the direct counting method was applied to each segment individually.
This approach, referred to as partitioned direct counting, results in a wide spread of the determined rates, from 0.69 to 1.89\,s$^{-1}$ with a standard deviation of 0.28\,s$^{-1}$ for 1.00\,V.
For 0.95\,V, the rates range from 0.08 to 0.60\,s$^{-1}$ leading to a standard deviation of 0.13\,s$^{-1}$.
The rates determined using the direct counting method therefore depend on the sample size.
As a result, assessing whether a given sample size is sufficient is not straightforward.

The WTD method, as discussed in the main text and suggested by its name, consists in determining the distribution of the waiting times.
Spurious and missed switching events primarily affect the distribution at very short waiting times, while the absence of long waiting times---due to limited experimental sample size---impacts the corresponding tail of the distribution.
However, the influence of such erroneous points is limited, as the switching rates are ultimately extracted from a fit to the distribution.
In addition, inspecting the distribution—such as verifying that it follows an exponential decay—provides a means to assess whether the sample size is sufficient.

Several technical aspects of the fitting procedure deserve mention.
First, we systematically excluded the first point of the WTD from the fit, as it is often affected by detection errors.
Second, instead of fitting an exponential function directly to the WTD, we performed a linear fit to the logarithm of the WTD.
This choice mitigates the problem that standard fitting routines tend to weigh high-count data points more heavily than low-count ones, which can bias the fit.
Fitting in logarithmic space significantly reduces this imbalance.
Third, in principle, the distribution should be normalized to one, which imposes a constraint on the offset of the linear fit.
However, we performed the fits without constraining the $y$-intercept, as spurious and missed events could strongly affect the normalization and thereby introduce systematic errors.

Applying the WTD method to the same dataset yields rates of 1.19 and 0.23\,s$^{-1}$ for 1.00 and 0.95\,V, respectively.
For the reasons outlined above, the rates extracted using this method are expected to be significantly less affected by systematic errors compared to those obtained via the direct counting method.
The obtained rates are 7\% and 17\% lower than those determined using the direct counting method for 1.00\,V and 0.95\,V, respectively.
This provides an estimate of the extent to which the direct counting method overestimates the switching rates, based on sample sizes of 444 and 204 switching events for 1.00 and 0.95\,V, respectively.
The discrepancy between the two methods may depend on several factors, including the noise level of the data, the bandwidth of the current amplifier, and the robustness of the algorithm used to distinguish genuine switching events from spurious noise.

Overall, this makes the WTD approach the most reliable method for filtering out potential measurement artifacts and statistical biases arising from a limited number of switching events.
We consider it therefore better suited for determining switching rates.

\section{Additional data sets at higher sample voltages}
\label{app:additionalData}
\begin{figure}[t]
	\includegraphics[keepaspectratio,width=0.45\textwidth]{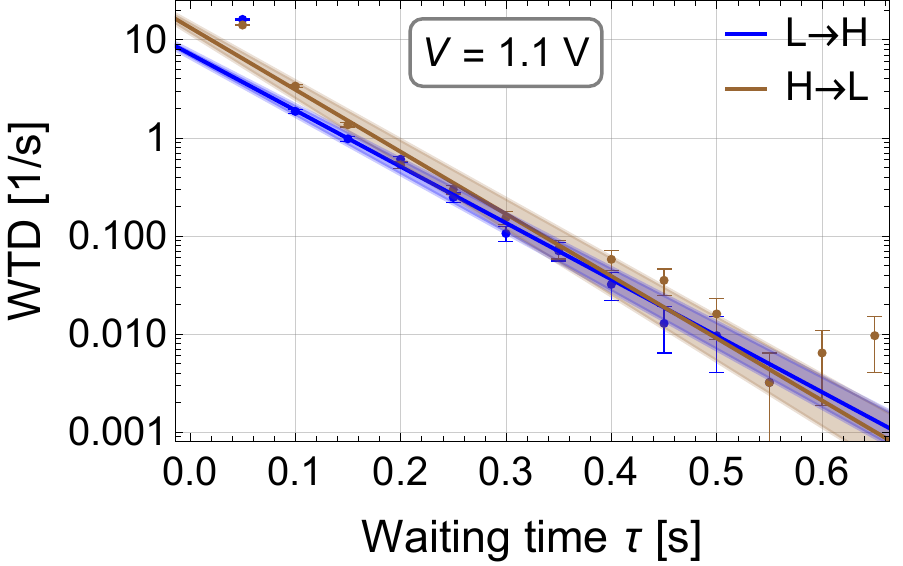}
	\par\vspace{0.5cm}
	\includegraphics[keepaspectratio,width=0.45\textwidth]{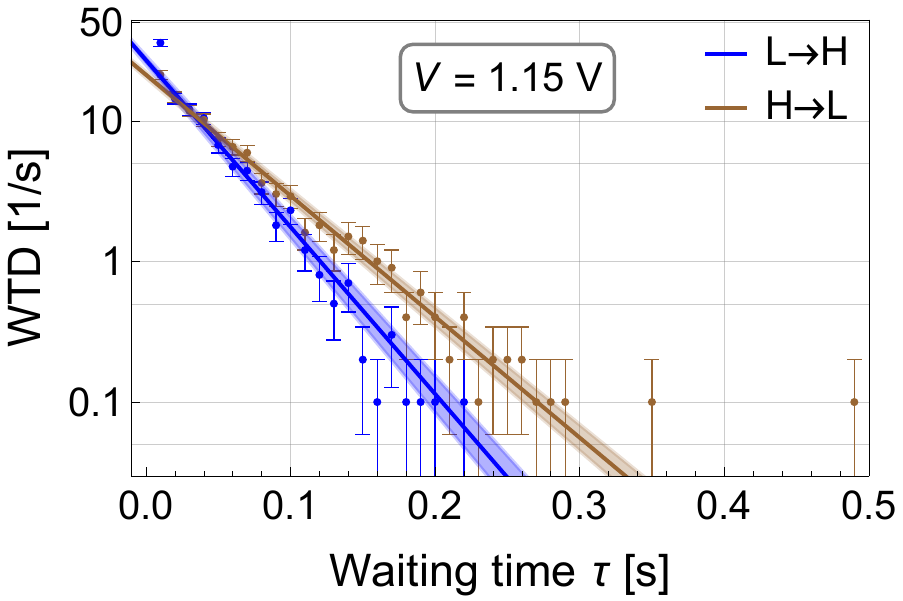}
	\caption{\label{fig:WTDAdditionalData} WTDs for sample voltages ${V=\qty{1.1}{V}}$ and ${V=\qty{1.15}{V}}$. The first data point lies above the fit for each case and is left out in the fitting process.}
\end{figure}

As mentioned in the main text, datasets acquired at sample voltages of $V=1.10\,\text V$ and $V=\qty{1.15}{V}$ may be affected by the limited bandwidth of current detection, particularly at short waiting times.
Nevertheless, the corresponding WTDs remain robust and display the expected exponential decay characteristic of the stochastic behavior inherent to the investigated system.
Minor deviations are observed at short delays (see, for example, the data point at 0.05\,s in the upper panel of Fig.~\ref{fig:WTDAdditionalData}), which can be excluded from the fit used to extract the switching rates.
As a result, the switching rates obtained from these datasets are reliable and are included in the analysis presented in the main text.

\section{Details regarding the fits}

\label{app:fits}
\begin{table}[h!]
    \centering
    \begin{tabular}{l|l|l}
         Rates & Value [s$^{-1}$] & Uncertainty [s$^{-1}$] \\
         \hline
         $W_{L^-\to H^-}$ &  $1.12\times 10^9$ & $2.86\times 10^{10}$ \\
        $W_{L^-\to L}$ & $5.70\times 10^{13}$ & $8.51\times 10^{13}$ \\
        $W_{H^-\to L^-}$ & $6.76\times 10^7$ & $9.88\times 10^{7}$ \\
        $W_{H^-\to H}$ & $4.90\times 10^{13}$ & $3.65\times 10^{13}$ \\
    \end{tabular}
    \caption{Rates and corresponding uncertainties inferred from the fits shown in Fig.~\ref{fig:Ergebnis}.}
    \label{tab:fitParam}
\end{table}
The dependence of the switching rate on the sample voltage,
%The evolutions of the switching rate with the sample voltage, 
inferred from the experimental data, are fitted with Eq.~\eqref{FitfunktionArctan}.
The rates, inferred from fits shown in Fig.~\ref{fig:Ergebnis}, are given in the Table~\ref{tab:fitParam}.
Owing to the large number of parameters, the uncertainties associated to the fitted rates are larger than or of the same order of magnitude as the rates themselves, which prohibits further consideration of these parameters.
The fits also yield the factors ${\alpha_L=(17.28\pm 449.45)\,\text V^{-1}}$ and ${\alpha_H=(1.07\pm 1.81)\,\text V^{-1}}$, defined in Eq.~\eqref{DirectCurrent} to connect the current directly flowing from the tip to the substrate to the electronic coupling $\Gamma_\text{tip}$ between the tip and the molecule.
The associated uncertainties are, once more, too large to really exploit those values further.

\bibliography{Bib_short.bib}

\begin{thebibliography}{39}
\bibitem [{foo()}]{footnote1}%
  \BibitemOpen
  \href@noop {} {}\bibinfo {note} {In contrast to the WTD fits, the fit of the
  switching yields was performed without weighting, which is typically based on
  the uncertainties of the individual data points. Applying such weighting
  would disproportionately emphasize the low-voltage yields, whereas our model
  more accurately captures the yield behavior at higher voltages.}\BibitemShut
  {Stop}%
\end{thebibliography}

\end{document}